# Field and temperature scaling of the critical current density in commercial REBCO coated conductors

Carmine Senatore, Christian Barth, Marco Bonura, Miloslav Kulich, Giorgio Mondonico

Département de Physique de la Matière Quantique (DQMP) and Département de Physique Appliquée (GAP), University of Geneva, Geneva CH-1211, Switzerland

**Abstract**

Scaling relations describing the electromagnetic behaviour of coated conductors (CCs) greatly simplify the design of REBCO-based devices. The performance of REBCO CCs is strongly influenced by fabrication route, conductor architecture and materials, and these parameters vary from one manufacturer to the others. In the present work we have examined the critical surface for the current density, $J_c(T,B,\theta)$, of coated conductors from six different manufacturers: American Superconductor Co. (US), Bruker HTS GmbH (Germany), Fujikura Ltd. (Japan), SuNAM Co. Ltd. (Korea), SuperOx ZAO (Russia) and SuperPower Inc. (US). Electrical transport and magnetic measurements were performed at temperatures between 4.2 K and 77 K and in magnetic field up to 19 T. Experiments were conducted at three different orientations of the field with respect to the crystallographic c-axis of the REBCO layer, $\theta = 0°$, 45° and 90°, in order to probe the angular anisotropy of $J_c$. In spite of the large variability of CCs' performance, we show here that field and temperature dependences of $J_c$ at a given angle can be reproduced over wide ranges using a scaling relation based only on three parameters. Furthermore, we present and validate a new approach combining magnetic and transport measurements for the determination of the scaling parameters with minimal experimental effort.

## 1. Introduction

Second generation $REBa_2Cu_3O_{7-x}$ (REBCO, RE = rare earth) coated conductors are complex multilayer materials. Processing technology and tape architecture vary from one industrial manufacturer to the other, resulting in largely different electromagnetic, electromechanical and thermal properties [1-3].

Coated conductor (CC) development has been driven so far by the perspectives of applications in the electrical utility sector. The efforts of manufacturers have focused on the optimization of the critical currents in low fields (1-3 T) and at temperatures ranging between 50 and 77 K. Only in the last few years the interest for high field magnets based on REBCO coils started to emerge, even if research in HTS coils made with BSCCO tapes goes back to mid-90s [4, 5]. Ongoing projects include solenoidal magnets in the 30 T range [1, 6, 7], 20 T dipole magnets for particle accelerators [8], the DEMO design studies for fusion power plants [9]. However, there have been limited reports on the electrical transport properties of REBCO tapes at 4.2 K in magnetic fields above 15 T [10-12].

The intrinsic anisotropy of REBCO superconducting properties determines the angular anisotropy of the critical current, $I_c$. Beyond the challenge of winding a coil using a tape with large aspect ratio, the anisotropy of the critical current introduces an additional layer of complexity to the magnet design. $I_c$ at T = 4.2 K and B > 10 T is maximum when magnetic field is parallel to the tape surface (i.e. in the crystallographic ab-plane of the REBCO layer) and minimum when the field is normal (i.e. parallel to the c-axis). Therefore, the most critical part in a tape-wound solenoid is at the coil ends, where the radial component of the magnetic field – perpendicular to the tape surface – is high. However, the $I_c(\theta)$ curve, θ being the angle between the field and the crystallographic c-axis of the REBCO layer, may show multiple peaks as an effect of nanoparticle doping, with a second maximum when B is parallel to the c-axis. A detailed knowledge of the anisotropy of $I_c$ at the operating temperature and field is thus crucial for magnet design. It is worth mentioning that Long et al. [13] and, more recently, Hilton et al. [14] proposed practical expression to fit the experimental $I_c(\theta)$ dependence, even in the presence of multiple peaks.

The protection of REBCO-based coils represents another critical issue. The operating temperature margin is clearly much larger compared to LTS-based magnets. This results in a much slower normal zone propagation velocity that makes difficult the detection of quenches [15, 16]. A magnet quench can be triggered when the winding temperature locally exceeds the current sharing temperature, $T_{cs}$, i.e. the temperature at which the current in the tape is equal to the critical current. The characterization of the temperature dependence of $I_c$ between 4.2 K and $T_{cs}$ is thus necessary to consolidate the electromagnetic models simulating the coil behaviour in case of quenches.

To qualify REBCO coated conductors for operation in very high field magnets, magnet design must be carefully adapted to the specific $I_c(T,B,\theta)$ characteristics of the tapes provided by the various manufacturers. On the other hand, the large performance variability would make necessary to conduct extensive critical current measurements in magnetic fields up to 30 T at various orientations and with temperatures ranging between 4.2 K and $T_{cs}$ for each conductor batch, and this would be an extremely time-consuming process. Scaling relations describing the electromagnetic behaviour of coated conductors are therefore essential for the proper design of the superconducting device. Hence, it is crucial to identify a minimum number of experimental results necessary for a comprehensive characterization of the electrical transport properties of REBCO tapes as well as determine the extrapolations' error margins. This is the motivation of the present paper. In particular, we propose a simple expression allowing the reconstruction of the $I_c(T,B)$ surface for a given field orientation using only three parameters. The paper is structured as follows:

- We have examined the electrical properties of coated conductors from six industrial manufacturers. Details about the tapes are given in Sec. 2.
- The temperature and magnetic field dependences of the critical current density were determined by transport and inductive measurements, as described in Sec. 3.
- Results are reported in Sec. 4 and discussed in Sec. 5.
- Sec. 6 is devoted to the conclusions.

## 2. Characteristics of the investigated samples

In the present work we have investigated the electrical and magnetic properties of coated conductors from six different industrial manufacturers: American Superconductor Co. (US), Bruker HTS GmbH (Germany), Fujikura Ltd. (Japan), SuNAM Co. Ltd. (Korea), SuperOx ZAO (Russia) and SuperPower Inc. (US).

Commercial coated conductors rely on two common features: a biaxially textured template, consisting of a flexible metallic tape coated with a multifunctional oxide barrier, and an epitaxial REBCO layer. The textured template is created by either deforming the metal substrate with the Rolling Assisted Biaxially Textured Substrate technology (RABiTS) [17] or by texturing the buffer layers deposited on the metal substrate by the so-called Ion Beam Assisted Deposition (IBAD) [18] and by its variant, the Alternating Beam Assisted Deposition (ABAD) [19]. The epitaxial REBCO layer is grown either by chemical routes, such as metal organic deposition (MOD) [20] and metal organic chemical vapor deposition (MOCVD) [21], or by physical routes, such as pulsed laser deposition (PLD) [22-24] and reactive co-evaporation (RCE) [25, 26]. The deposition of the precursors and conversion of the precursors into REBCO can occur in a single step (*in situ* process), as for MOCVD and PLD, and or in two steps (*ex situ* process), as for MOD and RCE. Typically, a few-µm Ag layer for protection against the moisture from the environment and a Cu layer for thermal and electrical stabilization are added to complete the conductor.

The characteristics of the investigated coated conductors are summarized in table I.

**Table I – Fabrication process and technical data of the investigated REBCO tapes**

| Manufacturer | Technology | Substrate (Material/Thickness) | Cu stabilizer (Type/Thickness) | REBCO thickness | Dimensions (w × t) |
|---|---|---|---|---|---|
| American Superconductor | RABiTS/MOD | NiW / 75 µm | laminated / 50 µm per side | 0.8 µm | 4.8 × 0.20 mm$^2$ |
| Bruker HTS (BHTS) | ABAD/PLD | Stainless steel / 100 µm | electroplated / 25 µm per side | 3.8 µm | 4.1 × 0.15 mm$^2$ |
| Fujikura | IBAD/PLD | Hastelloy / 75 µm | laminated / 75 µm on 1 side | 2.0 µm | 5.1 × 0.16 mm$^2$ |
| SuNAM | IBAD/RCE | Hastelloy / 60 µm | electroplated / 20 µm per side | 1.4 µm | 4.0 × 0.11 mm$^2$ |
| SuperOx | IBAD/PLD | Hastelloy / 60 µm | electroplated / 10 µm per side | 1.2 µm | 4.0 × 0.09 mm$^2$ |
| SuperPower | IBAD/MOCVD | Hastelloy / 50 µm | electroplated / 20 µm per side | 1.4 µm | 4.0 × 0.10 mm$^2$ |

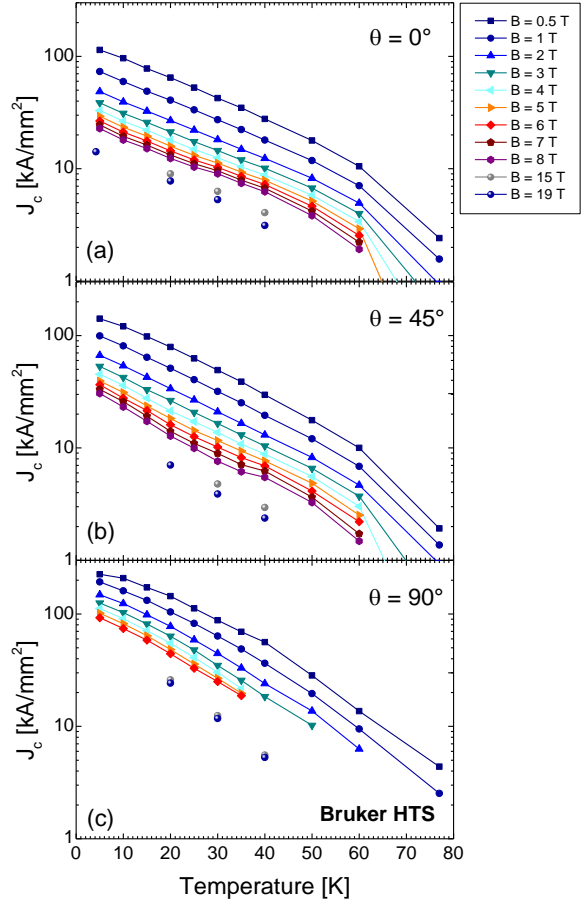

**Figure 1** Temperature dependence of the critical current density $J_c$ for the tape from Bruker HTS, measured for three orientations of the magnetic field, $\theta = 0°$ (a), $45°$ (b) and $90°$ (c). Orientations are referred to the angle between the magnetic field and the crystallographic c-axis of the REBCO layer. $J_c$ data are extracted from inductive ($B \leq 8$ T) and transport ($B > 8$ T) measurements.

## 3. Experimental details

Transport critical current and magnetization were measured over a broad range of temperatures and magnetic fields for three orientations of the tapes with respect to the field, $\theta = 0°$ and $\theta = 90°$, defined, respectively, when field is normal and parallel to tape surface, and $\theta = 45°$.

The characterization protocol of $I_c$ includes four temperatures, T = 4.2, 20, 30 and 40 K, in magnetic fields up to 19 T. In all following experiments, the critical current is defined as the current at the critical electric field of 0.1 µV/cm. Our setup for critical current measurements is limited to current values of 1000 A at 4.2 K and to ≈ 250 A above 4.2 K. Due to these limitations, we performed measurements on full width tapes only in the orientations with $\theta = 0$ and $45°$, while $I_c$ in parallel field ($\theta = 90°$) was measured on samples with reduced width. Two different techniques were adopted for the reduction of the REBCO layer width: chemical etching and electrical discharge machining (EDM). In the etching process, a lift-off photolithographic technique was used to obtain a 1 mm bridge at the center of the tape. This approach was preferred for coated conductors with electrodeposited stabilizer. EDM cutting was used for the tapes with laminated stabilizer: the sample width was reduced with a single cut from one side, varying the bridge width between 1 and 2 mm. In the case of reduced width samples, transport measurements were always repeated on different bridges in order to exclude etching or cutting defects.

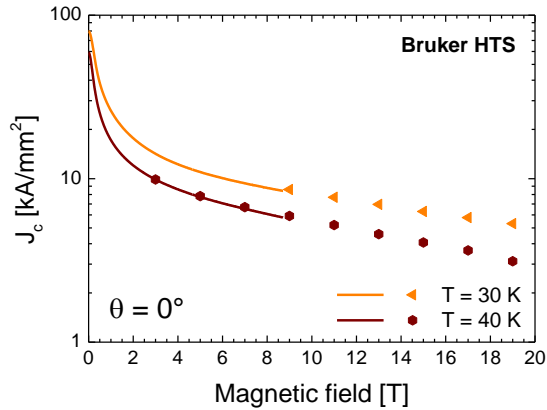

**Figure 2** Magnetic field dependence of the transport (full symbols) and inductive (solid lines) $J_c(\theta = 0°)$ at T = 30 K and 40 K for the Bruker HTS tape. The $J_c$ value at T = 40 K and B = 7 T from the transport $I_c$ was used to convert magnetization to critical current density.

Magnetization measurements in the field orientations with $\theta = 0°$ and 45° were performed using a vibrating sample magnetometer (VSM). Magnetization vs. magnetic field, M(B), loops were recorded at fixed temperatures between 4.2 and 77 K in fields up to 8.8 T. The field was swept at a rate of 1 T/min. At $\theta = 90°$, the signal from the superconductor is overwhelmed by the magnetic moment of the substrate. In fact, the area where the superconducting screening currents flow and thus the corresponding magnetic moment are small, being proportional to the REBCO film thickness. To overcome this problem, we prepared special samples by separating the superconductor from the substrate. It is well known that coated conductors are prone to delamination at the interface between REBCO layer and buffers. We exploited this structural weakness to peel off the stack made of Cu, Ag and REBCO layers from the substrate. A high sensitivity Superconducting Quantum Interference Device (SQUID) magnetometer was used to measure magnetization of the Cu/Ag/REBCO stack in parallel fields up to 7 T. As for the other orientations, magnetic field was swept at a rate of 1 T/min and temperature was varied between 4.2 and 77 K.

## 4. Experimental results

### *4.1. Temperature dependence of the critical current density*

To design high-field magnets based on REBCO tapes and which operate at 4.2 K it is necessary to parameterize the in-field critical current performance over a large range of temperatures, the temperature margin being of the order of 30 K. Figure 1 presents, in a lin-log chart, the temperature dependence of the critical current density in the REBCO layer, $J_c$, between 4.2 and 77 K for the tape from Bruker HTS. Magnetic field ranges between 0.5 and 19 T for the three orientations, $\theta = 0°$ (Fig. 1a), 45° (Fig. 1b) and 90° (Fig. 1c). Critical current density values extracted from the magnetization data were used to expand the explored window of temperatures and magnetic fields. Figure 2 compares the $J_c(B)$ curves at 30 K and 40 K as determined from transport and magnetic measurements. It is well known that the irreversible magnetization ($\Delta M$) is proportional to $J_c$ via a geometrical coefficient related to the length scale of the current flow. The value of $J_c$(40 K, 7 T) extracted from transport $I_c$ was used to set the coefficient of proportionality between ΔM and $J_c$ and thus to convert magnetization to critical current density. The two datasets, transport and inductive $J_c$, exhibit identical magnetic field dependence, as shown from the curves at T = 30 K and 40 K in Fig. 2.

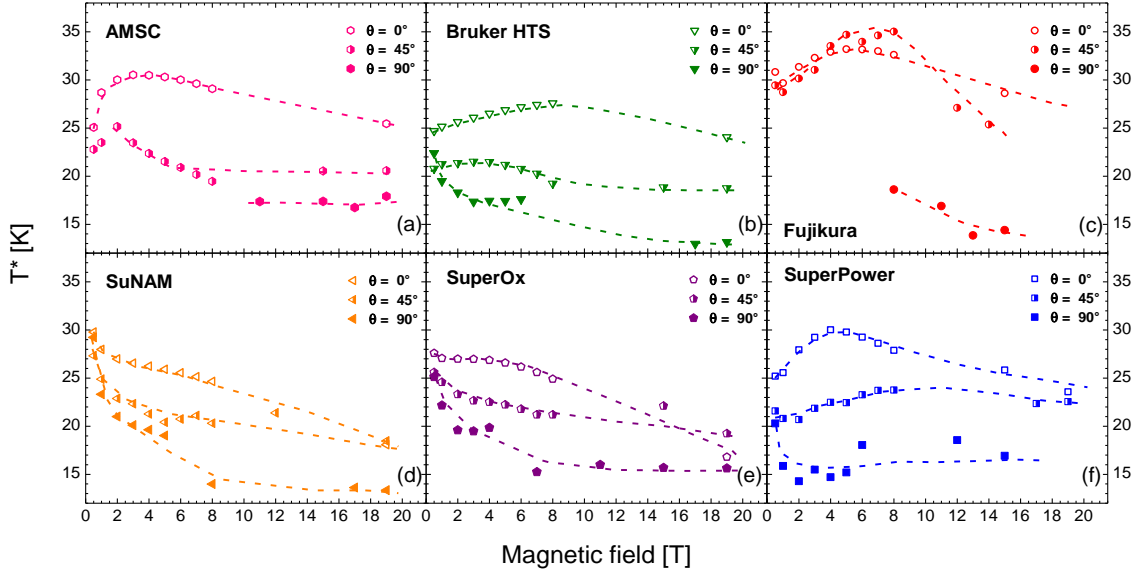

**Figure 3** Field dependence of the temperature scaling parameter, T*, as determined for the tapes from six industrial manufacturers: open symbols for θ = 0°, half-filled symbols for θ = 45° and solid symbols for θ = 90°. Dashed lines are given as guides to the eye.

From the curves shown in Fig. 1, it follows that the temperature dependence of $J_c$ can be described over a broad range of temperatures and magnetic fields by the equation

$$J_c(T,B) = J_c(T=0,B)e^{-T/T^*}. \qquad (1)$$

The exponential T-dependence holds up to ≈ 50 K for θ = 0° and 45°, and between 10 K and ≈ 40 K for θ = 90°, with a typical error below 2%. At θ = 90°, the higher is the field the lower is the deviation from the exponential behaviour at temperatures below 10 K. We encountered the same scaling behaviour for all the investigated tapes regardless of the manufacturer.

The temperature dependence of $J_c$ is determined by the thermal activation processes associated to the pinning centres. In particular, the exponential decay of $J_c$ is connected to the presence of defects generating weak isotropic pinning and T* is the characteristic pinning energy at these defects [27, 28]. A high value of T* implies a slow decrease of $J_c$ with T. The values of T* vary significantly from one manufacturer to the others. The dependence of T* on B may also vary from tape to tape of the same manufacturer, as the vortex pinning landscape depends on the fabrication route and on the possible addition of artificial precipitates.

T* values have been obtained by fitting the experimental results by Eq. 1 in the temperature range where the linearity in the lin-log plot is observed. Results are summarized in Figure 3 for the examined CCs. At the lowest field, B = 0.5 T, T* values span between 20 K and 30 K. At higher fields, the general trend is T*(0°) > T*(45°) > T*(90°). The only exception comes from the Fujikura CC, which exhibits T*(45°) > T*(0°) between 4 and 8 T.

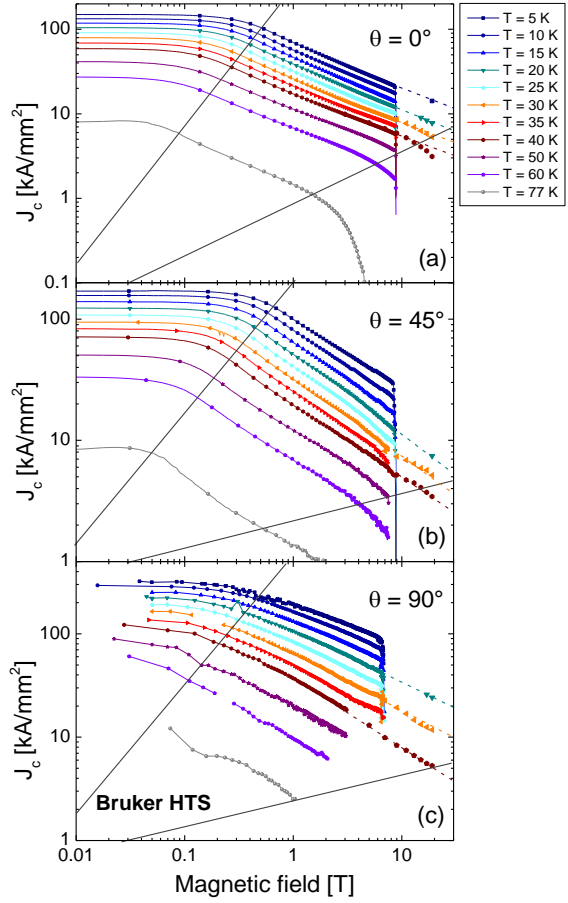

**Figure 4** Magnetic field dependence of $J_c$ between 5 K and 77 K for the tape from Bruker HTS. Measurements were performed for three orientations of the magnetic field, θ = 0° (a), 45°(b) and 90° (c). $J_c$ data are extracted from inductive (smaller symbols) and transport (larger symbols) measurements. Magnetic and transport data are in good agreement at all temperatures using the abovementioned normalization at 40 K, 7 T.

Tapes from SuNAM and SuperOx exhibit a monotonic decrease of T* with field independently of the orientation. For the other coated conductors, T*(0°) and T*(45°) reach a maximum between 2 T and ≈ 8 T followed by a smooth decrease as field increases, while T*(90°) decreases (BHTS and Fujikura) or stays nearly constant (AMSC and SuperPower) with increasing field.

The exponential T-dependence in eq. (1) provides a simple rule to estimate the variations of $J_c$ with temperature. If temperature is varied by ΔT, the value of $J_c$ changes by a factor $e^{-\Delta T/T^*}$ that does not depend on initial and final temperatures. For a given tape, the temperature lift-factor depends only on B.

### 4.2. *Magnetic field dependence of the critical current density*

Typical results for the magnetic field dependence of $J_c$ are reported in Figure 4, which shows exemplary data measured for the BHTS tape in the temperature range between 5 K and 77 K. The curves, obtained by electrical transport and inductive measurements, show a low field plateau followed by a smooth decrease of $J_c$ for fields higher than a given threshold. The threshold field depends on temperature and orientation. Its value is typically ≤ 0.5 T between 5 K and 77 K. The log-log plot of the data reveals that the field-induced decrease of $J_c$ is well described by a power law dependence, i.e. $J_c \propto B^{-\alpha}$, with an uncertainty of 5% or better. We observed the power law regime in the field range 0.5 – 19 T for T ≤ 30 K, but only between 0.1 and 1 T at 77 K. When approaching the irreversibility line, $J_c$ decreases faster and departs from the power law.

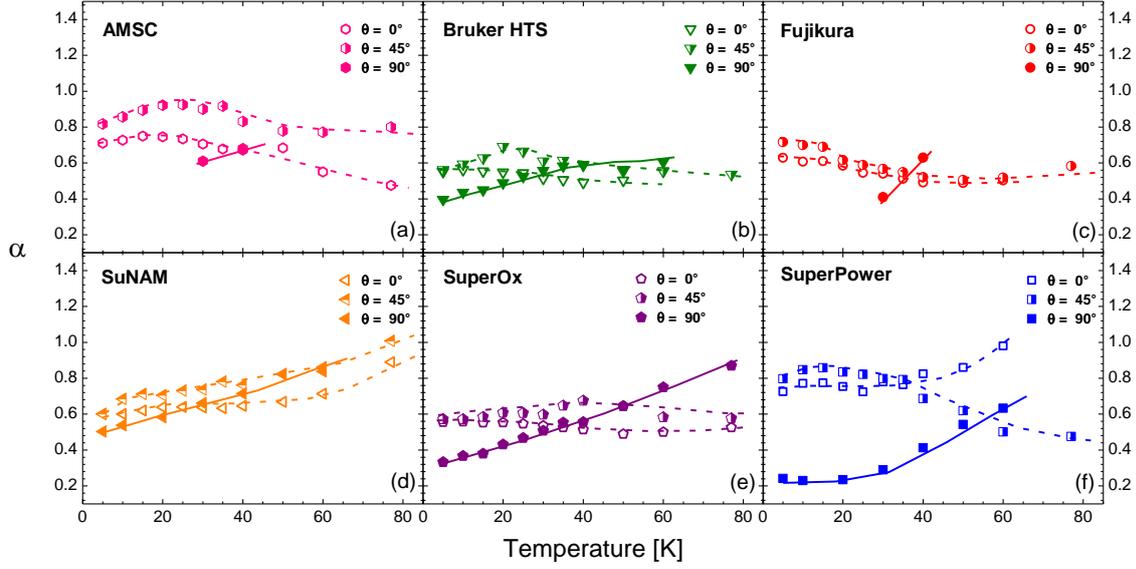

**Figure 5** Temperature dependence of the field scaling parameter, α, for $J_c$ as determined for the tapes from six industrial manufacturers: open symbols for θ = 0°, half-filled symbols for θ = 45° and solid symbols for θ = 90°. Dashed and solid lines are given as guides to the eye.

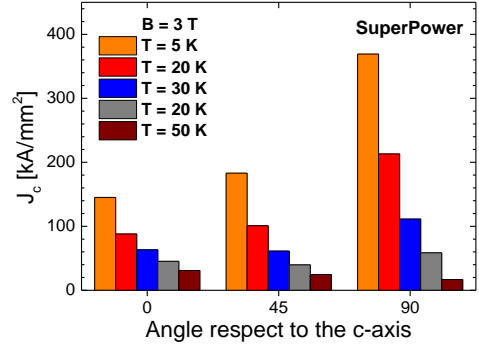

**Figure 6** $J_c$ at B = 3 T for the three angular orientations θ = 0°, 45° and 90° and the temperatures T = 5 K, 20 K, 30 K, 40 K and 50 K, as measured for the SuperPower tape. As a consequence of the artificial pinning, $J_c$(50 K, 3 T, 0°) > $J_c$(50 K, 3 T, 90°).

Figure 5 shows the temperature dependence of the power law coefficient α at θ = 0°, 45° and 90° for the investigated samples. α values have been obtained by fitting the $J_c$(B) curves using a power law in the field range where the linearity in the log-log plot is observed. α(0°) is almost constant in the temperature range 5 – 40 K, the lowest and the highest α-values being ≈ 0.55 for SuperOx and BHTS and ≈ 0.75 for SuperPower, respectively. The value of α(0°) is lower for the PLD-grown REBCO films (BHTS, Fujikura, SuperOx). From a fundamental point of view, the dependence with α = 0.5 is determined by the interaction between vortices nucleated at the grain boundaries and those strongly pinned to the dislocation cores [29, 30]. In the coated conductors where the REBCO layer is grown by chemical routes, dislocations do not connect from the bottom to the top of the film due to the grain boundary meandering [31, 32]. Therefore, c-axis correlated disorder becomes weaker and point-like defects play a role in the pinning landscape, leading to higher values of α [31].

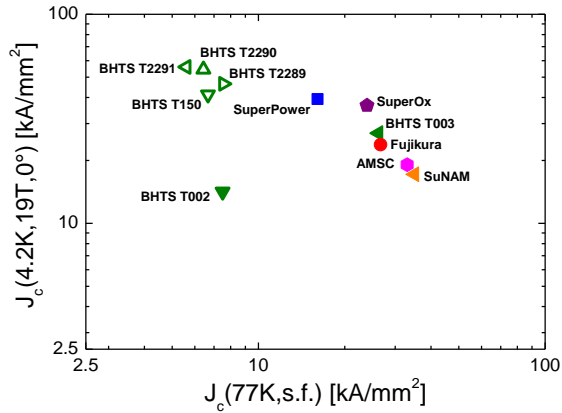

**Figure 7** $J_c$ of the REBCO layers measured at 4.2 K, 19 T, $\theta = 0°$ versus $J_c$ of the same layers measured at 77 K, self-field for six different manufacturers.

The higher is the value of α, the faster is the decrease of $J_c$ with increasing field. We observe that at temperatures below 30 K, α(90°) is smaller than α(0°) for all the examined tapes, while at higher temperatures the α(90°) curve intersects α(0°). The interpretation of this behaviour is intuitive. Since $J_c$ goes to zero at $T_c$ regardless of the field orientation, the higher is $J_c$ the faster has to be its in-field decrease when increasing temperature. In REBCO tapes without artificial pinning, the highest $J_c$ is observed for the orientation corresponding to $\theta = 90°$ and it follows that α(90°) has to become larger than α(0°) when temperature approaches $T_c$. The tape from SuperPower with artificial pinning represents a case apart. α(0°) stays higher than α(90°) over the explored temperature range and exhibits an upturn curvature at 40 K (see Fig. 5f). The manufacturer introduced BZO nanosized columns oriented along the REBCO c-axis to modify the pinning landscape and improve the electrical transport properties. One of the consequences is that the angular dependence of $J_c$ in the field region below 10 T varies with the temperature. This is illustrated in Fig. 6 where we report the values of $J_c$ for the three orientations at B = 3 T and T = 5, 20, 30, 40 and 50 K. Below 30 K, the orientation with $\theta = 0°$ exhibits the lowest critical current. In the range from 30 K to 40 K, the minimum $J_c$ was measured at $\theta = 45°$, while at T = 50 K we observe a peak at $\theta = 0°$, i.e. $J_c(0°) > J_c(45°) > J_c(90°)$. When approaching $T_c$, the in-field decrease has to be faster in the orientation corresponding to the highest $J_c$ that, in case of the SuperPower tape, corresponds to α(0°) > α(90°).

## 5. Discussion

Figure 7 reports a plot of $J_c$(77 K, self field) versus $J_c$(4.2 K, 19 T) measured at $\theta = 0°$ for various tapes, including the CCs listed in table I and five additional tapes from BHTS (T003, T150, T2289, T2290 and T2291). Our results show that it is not possible to find a univocal correlation between critical current values at low temperature/high field and at 77 K/self field when comparing the performance of CCs from various manufacturers or even from different production batches of a single manufacturer. A recent study on REBCO thin films made by MOCVD, both doped and undoped, finds a linear correlation between $J_c(T, B, \theta = 0°)$ and $J_c(77\ K, 3\ T, 0°)$ from 77 K down to 20 K and magnetic fields up to 9 T [33]. This is not the case for the commercial tapes examined in this work. In particular, we observe that tapes with inferior performance at 77 K yield superior critical currents at low temperature/high field. This occurs in the tapes from Bruker specially developed for high-field magnet applications, where the addition of nanoscale defects tailored in shape, size and spatial distribution leads to a strong enhancement of vortex pinning.

In spite of the large variability of CCs' performance, our analysis shows that field and temperature dependences of the critical current density at a given angle are well described by the simple expression

$$J_c(T, B) = J_c(T = 0, B = 0) e^{-T/T*} \cdot B^{-\alpha} \qquad (2)$$

over a wide range of temperatures and magnetic fields. In particular, magnetization data show that Eq. (2) holds for fields starting from ~0.1 T up to ~60% of the irreversibility field, $B_{irr}$, with a moderate dependence of α on temperature below 40 K. On the other hand, the scaling parameter for the temperature dependence, T*, was found to vary with magnetic field and the largest variations are encountered for the orientation with θ = 90°.

## 6. Conclusions

In summary, we have investigated the electromagnetic properties of REBCO coated conductors from six industrial manufacturers. The analysis proposed in this paper shows that the critical surface $J_c(T, B)$ at a given orientation can be reconstructed with a reduced number of measurements by combining the results of transport and magnetic measurements. To this end, it is sufficient to

- measure the transport $I_c$ at a single temperature over a field range that partly overlaps the range explored by magnetization.
- use the values of $J_c$ extracted from the transport $I_c$ to set the coefficient of proportionality between irreversible magnetization and critical current density.

Thanks to the lower technical complexity of magnetization measurement, it becomes easy to explore the critical surface over a wide range of fields and temperatures. Our study provides also the scaling relations for the field and temperature dependences of $J_c$. The experiments show that, in spite of the large variability of the CCs' performance, the $J_c$ of the examined tapes follows an exponential law for the temperature dependence, between 4.2 K and 50 K, and a power law for the field dependence, up to 60% of $B_{irr}$. Moreover, the temperature and field scaling parameters can be determined with minimal experimental effort by combining magnetic and transport measurements. This analysis is intended to provide a basic tool to magnet designers for modelling the behaviour of a high field insert.

**Acknowledgments**

Financial support was provided by the Swiss National Science Foundation (Grant No. PP00P2_144673 and Grant No. 51NF40-144613). Research also supported by FP7 EuCARD-2 http://eucard2.web.cern.ch. EuCARD-2 is cofounded by the partners and the European Commission under Capacities 7th Framework Programme, Grant Agreement 312453. The authors warmly acknowledge Damien Zurmuehle for his technical assistance.